# Information Seeking and Communication among International Students on Reddit

Chaeeun Han[1], Sangpil Youm[2], Hojeong Yoo[3], and Sou Hyun Jang[4]

[1]Pennsylvania State University, University Park, PA, USA
[2]University of Florida, Gainesville, FL, USA
[3]University of Michigan, Ann Arbor, MI, USA
[4]Korea University, Seoul, Korea

## Extended Abstract

International students have increased over 40 years from 1980s (Bound, Braga, Khanna, & Turner, 2021). However, the number of international students has dropped by 15% during the COVID-19 (Gewin, 2022). Throughout this pandemic, the vulnerability of international students has been magnified by issues such as perceived discrimination, stringent border regulations, and limited information access (Zhang, Hsu, Fleming, Liu, & Hahm, 2023; Chen, Li, Wu, & Tong, 2020; Firang, 2020). Amidst the uncertainties of the COVID-19 pandemic, there has been an increased demand for diverse information, with college students commonly relying on the Internet and social media platforms as their primary sources (Olaimat, Aolymat, Shahbaz, & Holley, 2020). However, few studies focused on international students' difficulties, their unmet information needs, and their information seeking behavior in online community. To address this gap in the literature, our current study aims to achieve the following objectives.

First, by analyzing questions posted on Reddit, one of the most popular social media platform, we seek to compare the changes in the topic of information sought by international students before and during the COVID-19 pandemic. Second, we examine the recurrence of specific topics and assess whether the number and content of recurring questions differ before and during the COVID-19 pandemic. Last, we explore the attributes of recurring questions, with a focus on determining whether these inquiries were resolved through communication with other international students on the Reddit.

In this study, we choose Reddit since it provides adequate online channels and public settings to discuss specific topics, as well as insights into information seeking and communication among its users (Proferes, Jones, Gilbert, Fiesler, & Zimmer, 2021). We explore posts and comments within the subreddit *r/f1visa*, which was created in 2014 with approximately 18,000 members, emerge as a focal point for communication regarding F1 Visa status inquiries. The dataset spans from August 20, 2015, to December 31, 2022, encompassing a total of 8,006 posts and 33,981 corresponding comments. We categorize the data into two periods: pre- and during-COVID. This division allows us to examine the impact of COVID-19 on the behavior of users in this subreddit. Statistical analyses indicate changes in user engagement patterns and participation levels, with notable increases observed in during-COVID posts and comments as shown in Table 1.

To discern the underlying topics within the subreddit discussions, we applies the BERTopic model, utilizing pre-trained language embeddings and custom class-based TF-IDF clustering. We employ a coherence score to determine the optimal number of topics, selecting six as the most coherent representation. Formulating representative topic name involves leveraging Large Language Model (LLM), ChatGPT, thereby enhancing interpretabliity of the extracted topics by facilitating Human-in-the-loop approach. We also analyze the differences in topics between individuals who ask a question only once and those who pose recurring questions. Additionally, we assess the coherence of a user's recurring questions using cosine similarity, which evaluates the semantic coherence of user posts and provides insights into persistent topics and recurring inquiries.

First, there has been an increase in the frequency of users submitting multiple posts during COVID-19. This trend is supported by a decrease in the percentage of authors making only one post, dropping from approximately 88% before the pandemic to around 73% during it, as illustrated in Figure 1. In contrast, there remains a consistent pattern across both periods, with the majority of users consistently posting only one question. However, the rise in the proportion of users posting multiple questions during the pandemic, from 12% to 27%, underscores the need to delve deeper into the motivations and patterns of these users. Understanding the behavior of users who create several posts can provide valuable insights into their information-seeking strategies and challenges they face.

Second, as depicted in Figure 2, both in pre- and during-COVID-19 periods, all users who create multiple posts show non-negative cosine similarity, inferring a tendency to generate recurring questions. We measure the cosine similarity across individual users' posts to assess the coherence of posts per user for investigating the traits of multiple posts. While the recurring tendency present in both periods, prevalent topics diverge between each period. In pre-pandemic, predominant topics revolved around job opportunities and visa-related inquiries. Conversely, during-COVID-19 topics expanded to financial interviews, academic preparations and tax-related issues. Prevalent topic among recurring questions and noteworthy changes during COVID-19 are illustrated in Table 2.

Third, we observe that international students who had recurrent posts tend to revisit their own questions more frequently, as measured by the frequency of the asking user's appearance in the comments of their own post. This behavior suggests that recurring asking questions users are more actively involved in communication with other users. To understand their interactive communication pattern, we analyze the word frequency of the comments in their questions, visually represented in a word cloud (Figure 3). Comments from single-question asking user are characterized by expressions of gratitude (e.g., *Thank*), whereas those from recurring questions users commonly includes words (e.g., *need* or *still*), indicative of persistent information seeking behavior.

Consistent with prior research by (Sin, 2015), our study supports the observation that international students prioritize seeking information related to legal matters and employment over academic resources. However, before and during the COVID-19 pandemic, there has been a shift in the focus of information-seeking among international students. Pre-pandemic, topics mainly centered around job applications in fields like data science and law, while during-pandemic, recurring questions shifted to travel concerns, financial difficulties, and funding. Additionally, there was a change in the nuances of common topics, such as increased inquiries about passport refusals. These findings have policy implications, including the need for cooperation among universities, the U.S., and home countries to provide timely information access. Furthermore, there should be increased support for distributing timely and sufficient information in response to the recurring questions to meet the unfulfilled information needs of interna-tional students.

Based on our findings, future research should explore and compare types of information sought by international students across various platforms, including traditional outlets and social media like Facebook and Quora. Continuous monitoring of online platforms like Reddit is essential to protect international students from scams and misinformation, particularly given their heightened anxiety about future careers and visa statuses.

# Figures and Tables

Table 1: *r/f1visa* data overview

| | All Period | Pre-COVID | Post-COVID |
|---|---|---|---|
| Duration | 2015/08/20-2022/12/31 | 2015/08/20-2019/12/31 | 2020/01/01-2022/12/31 |
| # of posts | 8,006 | 534 | 7,472 |
| # of comments | 33,981 | 1,325 | 32,656 |
| Comments per post | 4.24 | 2.48 | 4.37 |
| # of post authors | 3,814 | 325 | 3,519 |
| # of comment authors | 5,197 | 326 | 4,941 |

Table 2: Topic among posts of individuals created more than two. Note: Given examples have been rephrased to ensure anonymity and privacy.

| | Topic | Keywords | Example |
|---|---|---|---|
| **Pre-COVID 19** | | | |
| 1 | USCIS EAD Application: Processing Time, Card Receipt, and Start Date | ead, applications, card | I'm currently waiting on my new EAD card. Notice says my current card is extended 180 days, but my license tied to it expires this month. Tried renewing at the DMV, but they said the 180-day notice is not enough. Does anybody have any advice? |
| 2 | Data Scientist and Analyst Jobs: Entry-level Openings in the USA | data, jobs, scientist, data analyst | I applied to various roles in data-related fields, customized resume for 300 job applications with cover letters. I received 18 calls, but 12 rejected due to sponsorship needs. Despite good interviews, faced rejections, feeling frustrated with lack of support from companies. |
| 3 | Navigating B1/B2 Visa Interviews: Passport Requirements and Waiver Information | interview, b1, passport, waiver, b2 | On a B1 visa, I married a US green card holder. My parents-in-law offered to sponsor my legal education. (...) Despite delays in my visa extension and change of status application, I am aiming to switch to F1 status for the spring semester. What are my chances of changing from B1 to an F1 visa successfully? |
| 4 | Web Developer Jobs: Entry-level Opportunities in the USA | developer, entry level, jobs, web | I graduated in June and have been studying web developer. (...) Proficient in HTML, CSS, JavaScript, and Python, I wonder if these skills are enough for entry-level developer and if there are specific qualifications needed. |
| 5 | CPT Jobs in the USA: Requirements and Search Strategies | cpt, jobs, requirements, usa | My OPT ends Feb 11, 2019. Current company wants me to stay but I prefer not to. If my MS Finance degree gets STEM designation approved, can I apply for STEM extension even if I leave the company after the contract? |
| 6 | Entry-level Attorney and Analyst Jobs: Opportunities in the USA | entry level, attorney, jobs, opportunities | Discover entry-level attorney job openings in New York, USA, that enhance career prospects. |
| | **During-COVID 19** | | |
| 1 | Mastering the Financial Interview: Providing Proof of Funds of University Admission | interview, bank, funds, loan, financial | I am using a loan sanction letter from a private bank for faster I20 processing, planning to switch to a loan from a public bank for the visa interview. Is this acceptable to have different loan sources in the I20 and visa interview? |
| 2 | Optimizing STEM Internships: A Guide to CPT and Working Hours | cpt, time, company, internship, stem, degree, major | (...) Due to an oversight in my timesheet, I worked 22 hours one week without realizing it, and it was approved. Should I be worried? |
| 3 | Preparing for a New Academic Year: Essential Travel Tips for International Students | new, travel, country, home, classes, sevis, year | Due to Covid, I haven't visited my home country since arriving in the US in January 2021. My family advises against travel due to Covid concerns. Will this affect my F1 visa renewal for my master's program? |
| 4 | Crucial Steps in EAD Application: A Comprehensive Guide for STEM Students | ead, uscis, date, dso, application, card | Following USCIS's announcement on Covid flexibilities, OPT applications now receive full recommended OPT time. My program ended on May 2nd, 2020, but my EAD is valid until July 2nd, 2021. I seek a correction to extend it to August 20th, 2021 to recover lost authorization, but I am concerned about remaining EAD time and STEM-OPT risks. |

| 5 | Navigating IRS Requirements: Essential Tips for Filling Your Taxes | tax, taxes, resident, irs, insurance, non resident | As an international student considered a non-resident for tax purposes, I worked for my school's employer and earned gains from stock trading. Can someone clarify which forms I need to file this year and how to pay taxes on capital gains to the IRS? |
|---|---|---|---|
| 6 | Common Reasons for Passport Refusal and How to Address them | interview, embassy, administrative, processing, passport, documents | Has anyone successfully obtained an F-1 visa after being subjected to administrative processing during the US government shutdown? If so, what was the duration of the process? Are there any steps I can take to expedite the process or receive an estimated timeframe? |

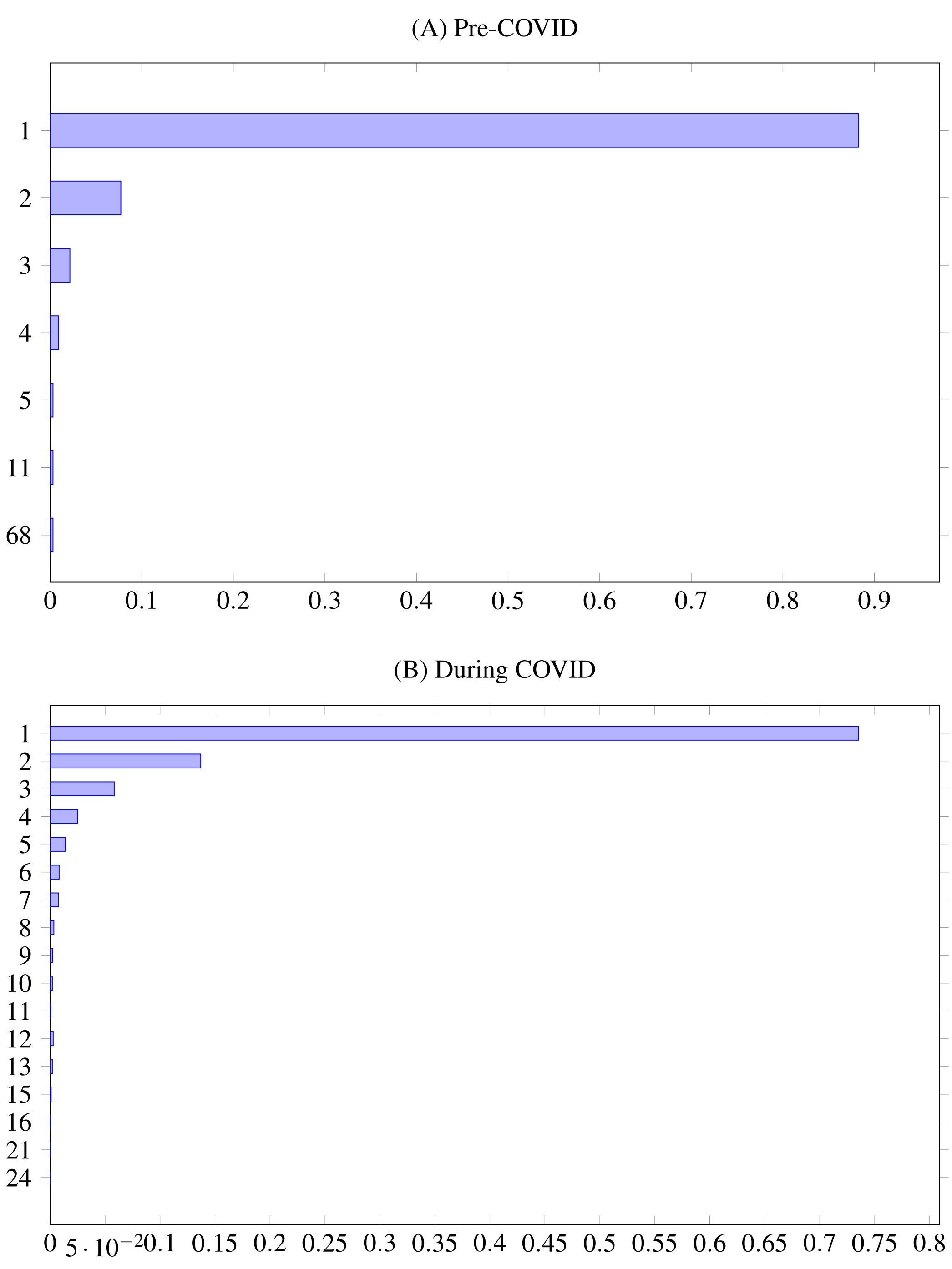

Figure 1: Number of Questions in Prior-and During-COVID. (A) Histogram illustrates the distribution of posts authored by individuals **prior to the COVID-19**. (B) Histogram illustrates the distribution of posts authored by individuals **during the COVID-19**.

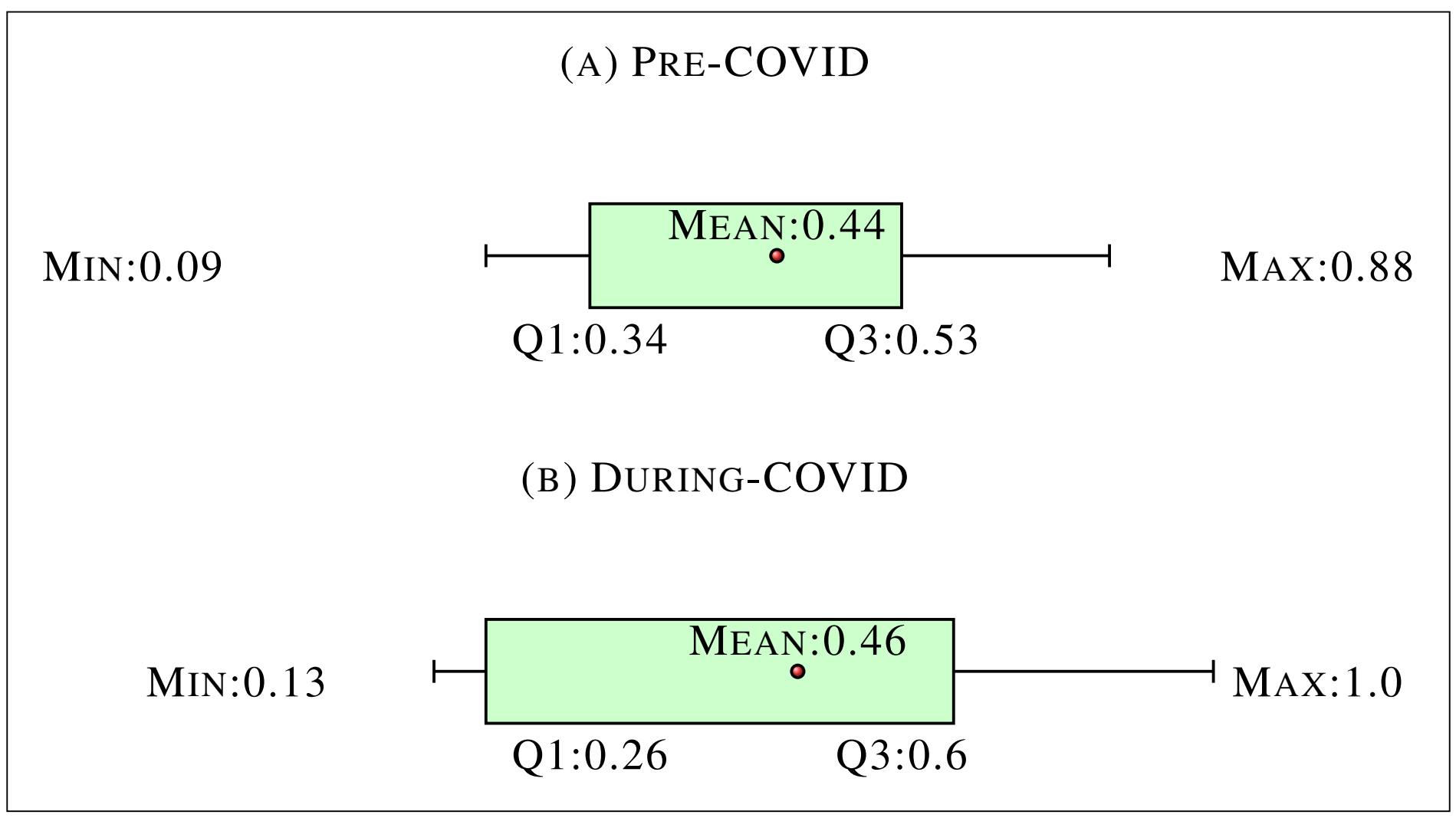

Figure 2: Box plots of cosine similarity. (A) Box plot illustrates the cosine similarities among posts from individual users posted more than twice **prior to the COVID-19.** (B) Box plot illustrates the cosine similarities among posts from individual users posted more than twice **during the COVID-19.**

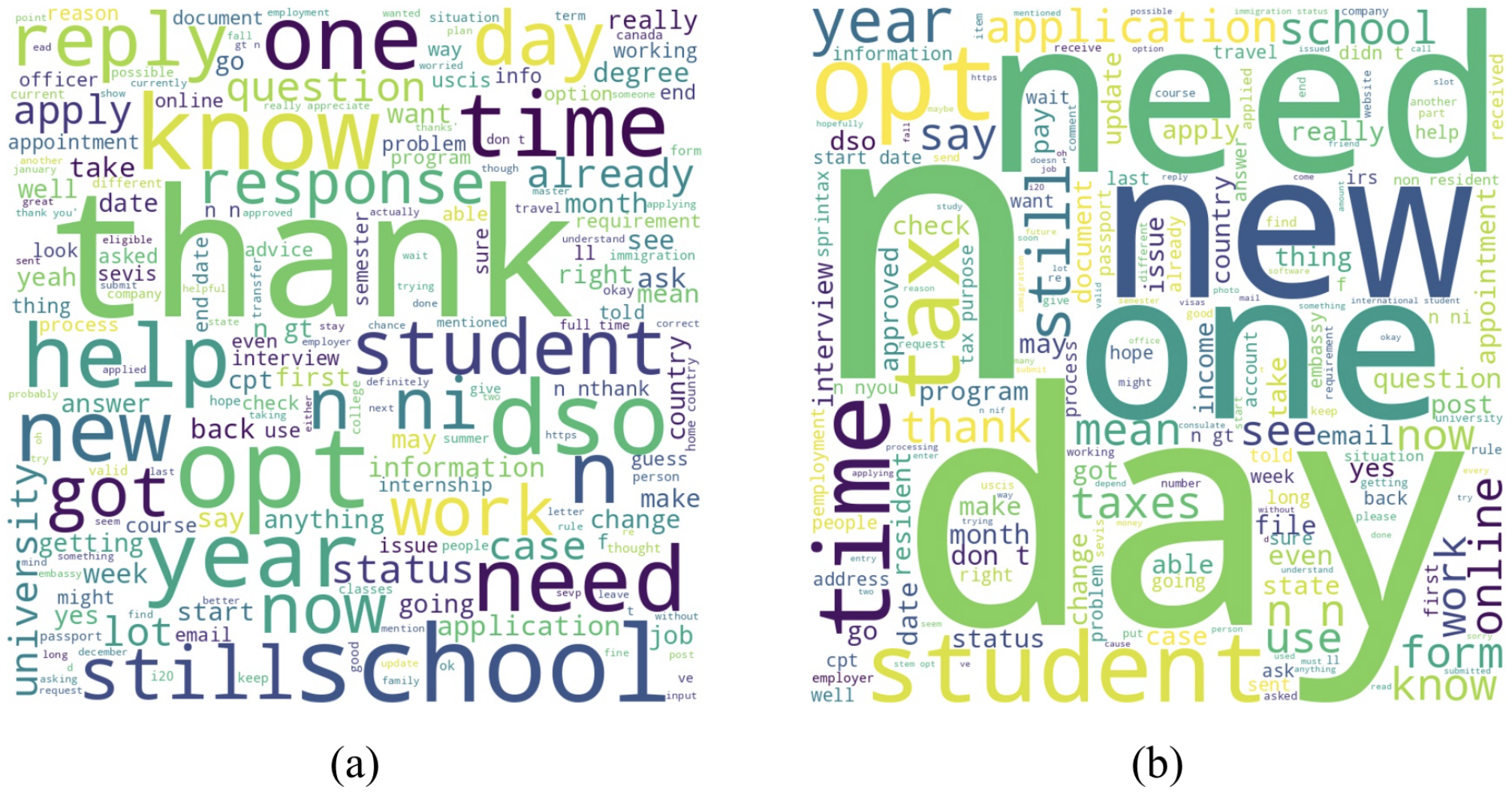

Figure 3: Word Cloud of Comments: (a) It analyzed comments from posts created by users who only posted one post. (b) It analyzed comments from posts created by authors who posted more than five posts.